# Development of a Platform to Enable Real Time, Non-disruptive Testing and Early Fault Detection of Critical High Voltage Transformers and Switchgears in High Speed-rail

Jiawei Fan, Ming Zhu, Yingtao Jiang, Hualiang Teng

Department of Electrical and Computer Engineering University of Nevada, Las Vegas

*Abstract*— **Partial discharge (PD) incidents can occur in critical components of high-speed rail electric systems, such as transformers and switchgears, due to localized insulation defects that cannot withstand electric stress, leading to potential flashovers. These incidents can escalate over time, resulting in breakdowns, downtime, and safety risks. Fortunately, PD activities emit radio frequency (RF) signals, allowing for the development of a hardware platform for real-time, non-invasive PD detection and monitoring. The system uses an RF antenna and high-speed data acquisition to scan signals across a configurable frequency range (100MHz to 3GHz), utilizing intermediate frequency modulation and sliding frequency windows for detailed analysis. When signals exceed a threshold, the system records the events, capturing both raw signal data and spectrum snapshots. Real-time data is streamed to a cloud server, offering remote access through a dedicated smartphone application, enabling maintenance teams to monitor and respond promptly. Laboratory testing has confirmed the system's ability to accurately capture RF signals and provide real-time PD monitoring, enhancing the reliability and safety of high-speed rail infrastructure.**

## I. INTRODUCTION

High-speed rail (HSR) has revolutionized transportation by providing rapid, efficient, and reliable services that connect major cities and regions. The inception of HSR dates back to the 1960s with the launch of Japan's Shinkansen, also known as the "bullet train," which set the benchmark for future developments in the industry (Givoni, 2006). Since then, numerous countries, including France, China, Germany, and Spain, have developed extensive HSR networks, showcasing significant advancements in speed, safety, and passenger comfort (Campos and de Rus, 2009). The California High-Speed Rail (CAHSR) project is also underway from Sacramento/San Francisco Bay Area to the Greater Los Angeles/San Diego (California, 2008; Fox et al., 2008).

The efficient and reliable operation of high-speed rail systems relies heavily on sophisticated high-voltage electrical facilities. Among these, transformers and switchgears play crucial roles in ensuring the seamless transmission and distribution of electrical power necessary for train operations. In HSR systems, transformers are essential for stepping up or stepping down voltage levels to meet the requirements of different stages in the power distribution process (Fig. 1. and Fig. 2.) (Sibal, 2010). They ensure that electrical power is efficiently transmitted from power stations to rail lines, maintaining the correct voltage levels needed for train operations. This step is vital for reducing energy losses and ensuring the smooth functioning of the rail network (Douglas et al., 2015). Switchgears are critical components that control, protect, and isolate electrical circuits within the rail system. They facilitate the safe operation of electrical circuits by managing the flow of electricity, enabling maintenance work, and protecting the system from faults. Switchgears ensure that power distribution remains uninterrupted and safe, even in the event of a fault, thereby maintaining the reliability and safety of high-speed rail operations (Feng et al., 2017). These components are designed to withstand significant electrical stresses and environmental conditions, necessitating robust design and meticulous maintenance. Their effective functioning is crucial for the continuous and safe operation of HSR, which in turn supports the demand for rapid and reliable transportation (IEC, 2014).

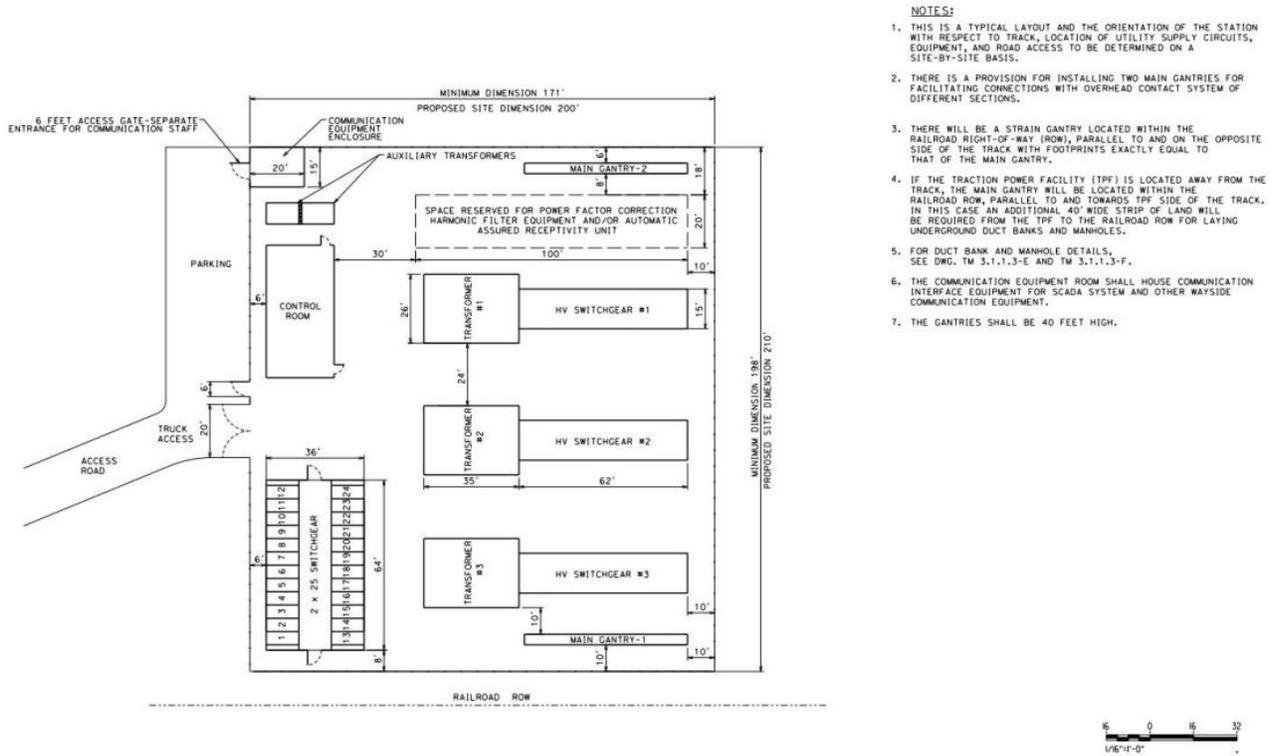

Fig. 1. Conceptual layout traction power substation with 3 high voltage transformers and switchgears

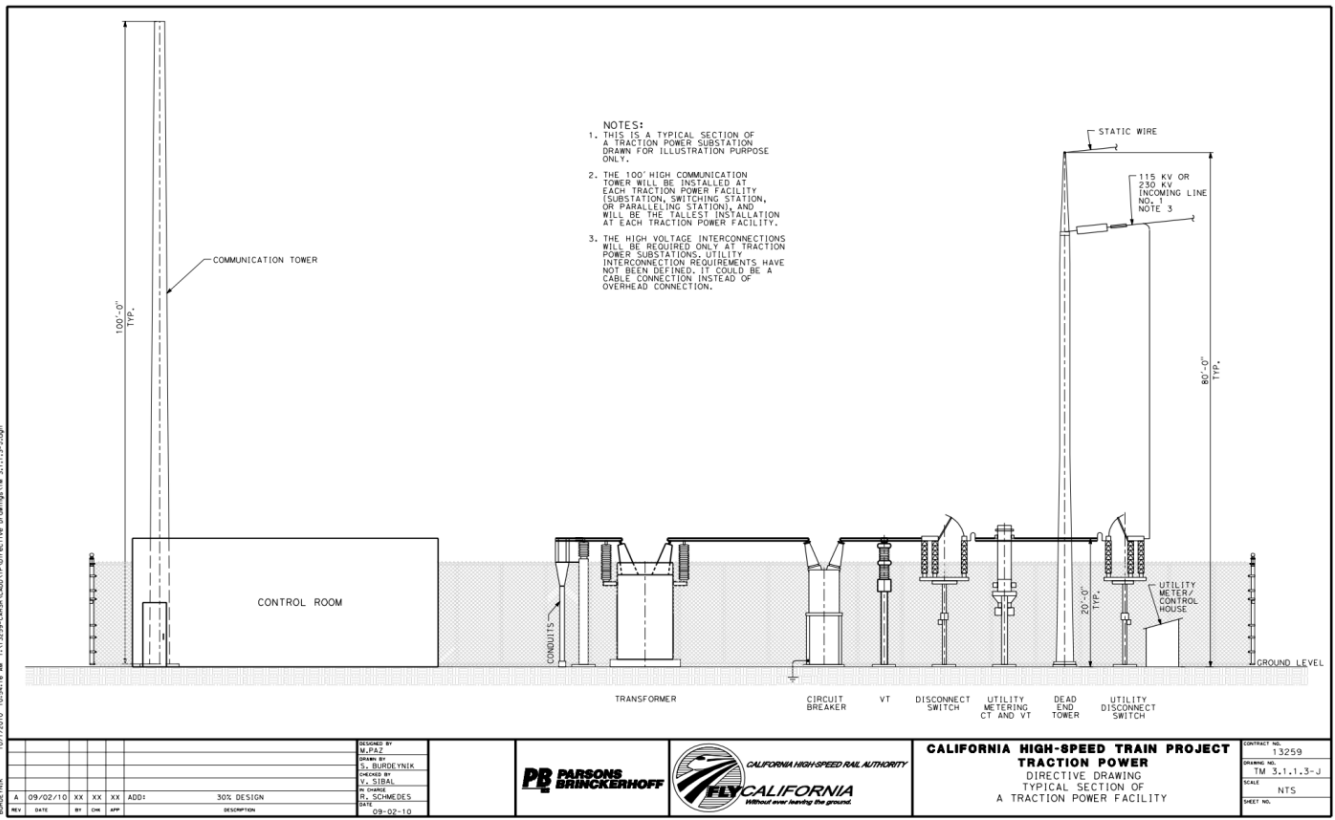

Fig. 2. Direct drawing of typical section of a traction power facility

However, these high-voltage assets are susceptible to partial discharge (PD), a phenomenon that can significantly undermine their performance and longevity. Partial discharge is a localized dielectric breakdown of a small portion of a solid or fluid electrical insulation system under high voltage stress. PD occurs when the electric field intensity exceeds the dielectric strength of the insulating material, causing localized discharges that do not completely bridge the insulation. These discharges typically result from defects such as voids, cracks, or contaminants within the insulation, which can arise during manufacturing, installation, or due to aging and environmental factors (F.H. Kreuger, 1989). The occurrence of PD in high-voltage transformers and switchgears can have severe consequences. PD generates heat and chemical byproducts that degrade the insulating material over time. This degradation leads to erosion, cracks, and carbonization, progressively weakening the insulation's ability to withstand high voltages (Fig. 3.). In transformers, PD can cause insulation failure, leading to costly repairs and significant downtime (Tenbohlen et al., 2017). In switchgears, PD can initiate surface tracking and arcing, which may result in equipment failures and pose fire hazards (Paoletti and Baier, 2001). According to the research (Davies, 2009; Hussain et al., 2016), 85% of high-voltage facility failures and asset damages are due to the PD activities (Fig. 4.).

To mitigate the risks associated with PD, it is crucial to implement effective monitoring and maintenance strategies. There are multiple approaches to detect PD, such as electromagnetic method (Jahangir et al., 2017; Judd et al., 2005a, 2005b), electrical method (Giussani et al., 2012; IEEE, 2024; Judd, 2011; Timperley, 1983), chemical (Ma et al., 2014; Sheng et al., 2004; Skelly, 2012; Sparkman et al., 2011), acoustic (Kim et al., 2017; Kučera et al., 2019; Qian et al., 2018), optical (Jürgen Fabian et al., 2014; Sarkar et al., 2015) and combinations of these methods (Hauschild and Lemke, 2019; Kim and Hikita, 2013; Kraetge et al., 2013; Yongfen et al., 2015). However, electrical and chemical methods require contact with the equipment, which may bring interference of normal railway operations; whereas acoustic and optical methods may be disrupted by the environmental noises that may affect the detection accuracy. On the other hand, since PD usually emits a radio frequency (RF) signal in a range between 300 and 1500MHz while there lack general background RF signals within this range, we can employ one or multiple ultra-high-frequency (UHF) antenna(s) to scan and detect RF signals in the air within this spectrum. With that, we can obtain a non-invasive/contactless non-disruptive real-time PD detection and monitoring system for high-voltage switchgears and transformers in high-speed rail infrastructure.

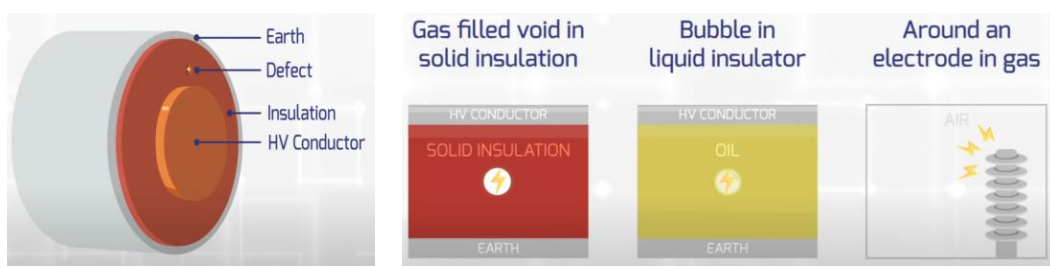

Fig. 3. Causes of PD in switchgears and transformers

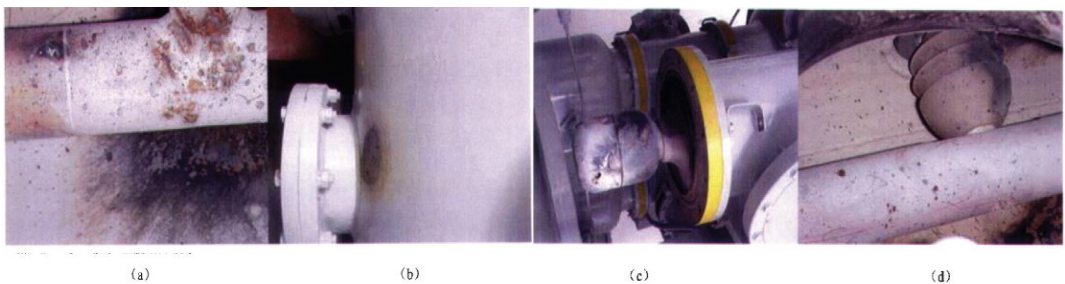

Fig. 4. Damages caused by PD hazards. a) inner surface burn, b) burn through the shield, c) contact damage, d) explosion of the pillar insulator

To continuously and losslessly monitor PD signals that span over a wide frequency range of 300MHz to 1500MHz, the data acquisition system (DAS) needs a minimum sampling rate of 3GHz (i.e., 3109 samples per second) or higher according to Nyquist Sampling Theory. Correspondingly, the DAS requires high-performance CPU and memory to save and even process this huge amount of data (i.e., approximately 3GB data per second even if using an 8-bit analog-to-digital converter for raw data). Developing such high-speed customized DAQ is usually costly and takes a long time to align all hardware to work effectively and reliably. In addition, a single UHF antenna may not suffice the job to constantly capture PD signals of such broadband, and employing multiple antennas will bring extra complexity to the circuit.

To resolve such challenges, we propose to use a single antenna to iteratively scan the spectrum from 100MHz to 2500MHz with intermediate frequency (IF) modulation (Bergmans, 2013; Whitaker, 2017). In this way, the system only has one input and is flexible to various scanning spectrums by adjusting the IF modulation frequencies. One drawback of this method is that the band-limited antenna can only scan a small range of RF spectrum at a time while leaving other spectrum unattended, causing the possibility of missing the PD. Nevertheless, as long as the PD events are repetitive and with proper configuration of the scanning rate, the system will catch PD occurrence confidently.

## II. DEVELOPMENT OF NON-DISRUPTIVE REAL-TIME PD DETECTION AND MONITORING SYSTEM

The hardware and software platform for PD signal capturing, detection and recording, as well as the cloud-based remote access on the smartphone.

### A. PD Data Acquisition Hardware Platform

We propose a compact and functionally comprehensive and accurate hardware platform as shown in Fig. 5. The system only consists of an antenna with a passing bandwidth of 800MHz and center frequency of 136MHz, a Tektronix RSA 306B real-time RF spectrum analyzer (Tektronix, 2015), and a computer (or other high-performance embedded system) with USB-3 interface.

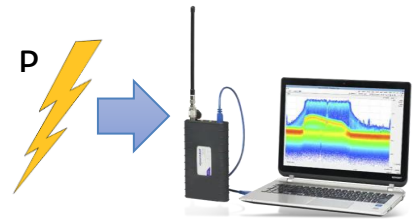

Fig. 5. PD signal/data acquisition system

### B. PD Data Acquisition Process

During each PD sampling process, the antenna contactlessly senses the PD signals in the air for a certain amount of time (i.e., sample duration, 10 milli-second in this study), and transmits the electrical signal to the RSA, which sets up a carrier wave to shift the ultra-high frequency PD signals to a lower spectrum (i.e., intermediate frequency modulation). The IF modulated signal will then be sampled, quantified and digitized by high-speed analog-to-digital converter (ADC), and the data stream will be transmitted to the computer via USB-3 connection (Fig. 6.). Here, we call data within one sample duration a data frame or sample frame, and each PD sensing/capturing will generate a frame. The computer reads the data frame via Tektronix application programing interface (API) (Tektronix, 2017), and calculate the data in both time domain in-phase/quadrature (I/Q) stream form (i.e., quadrature modulation) (Fig. 7.) (Gast, 2005) and spectrum form (Fig. 8.) using Discrete Fourier Transform (DFT) or Fast Fourier Transform (FFT), Eq. (1). The power is calculated as Eq. (2).

$$X[k] = \sum_{n=0}^{N-1} x(n) e^{-j2\pi kn/N} \quad (1)$$

$$Power = 10\,log\left(\frac{I^2 + Q^2}{1\,mW}\right) \quad (2)$$

If the system detects any signal power that exceeds a pre-set threshold (i.e., -50dB in this study) during each sample frame, the I/Q data and spectrum information of that frame will be recorded.

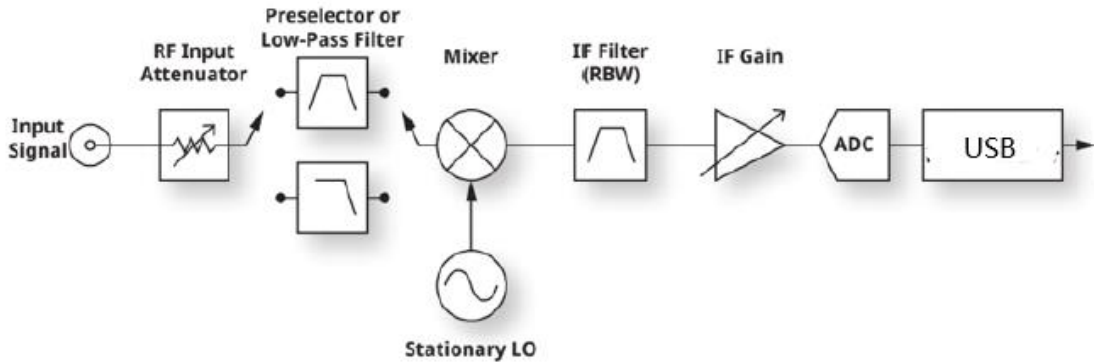

Fig. 6. Signal capturing, modulation and sampling inside the RSA (Tektronix, 2017)

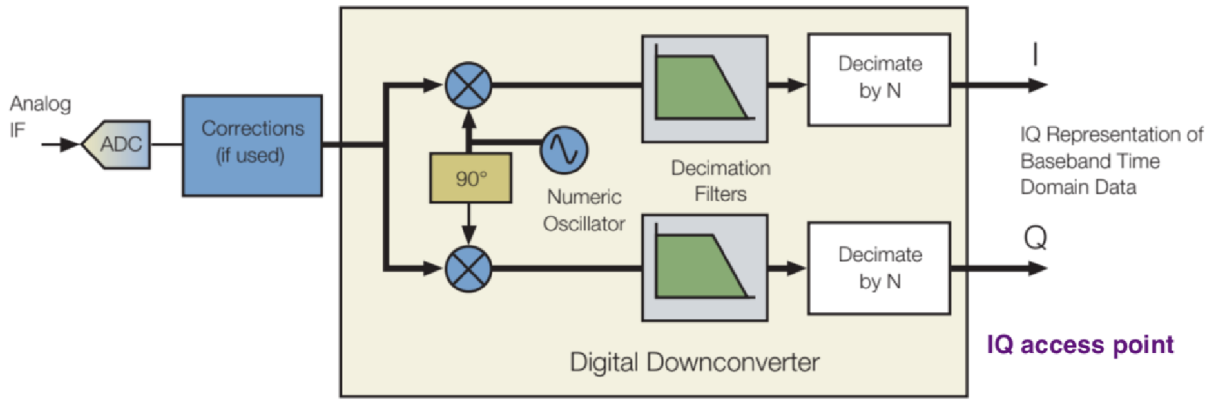

Fig. 7. I/Q modulation for data in time domain (Tektronix, 2017)

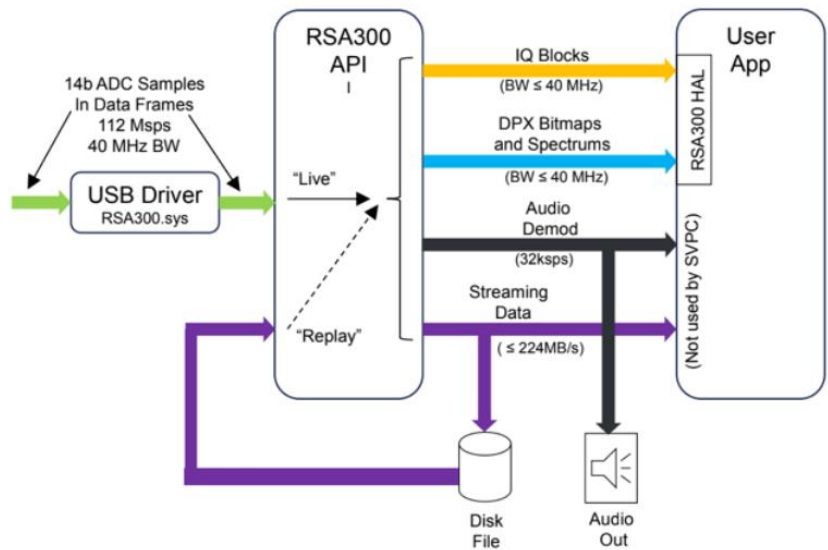

Fig. 8. RSA API and data flow in computer (Tektronix, 2017)

The API produces raw data files for IF and I/Q measurements under .r3f and .siq format (Fig. 9.), which can be further decoded to .csv file with time sequence but at a cost of much larger file size (Fig. 10.). The fundamental functions of data acquisition are fulfilled by C/C++ based on RSA 306B API, and a python interface is implemented on top of the C/C++ code and RSA API for the simplicity of user interaction and process control.

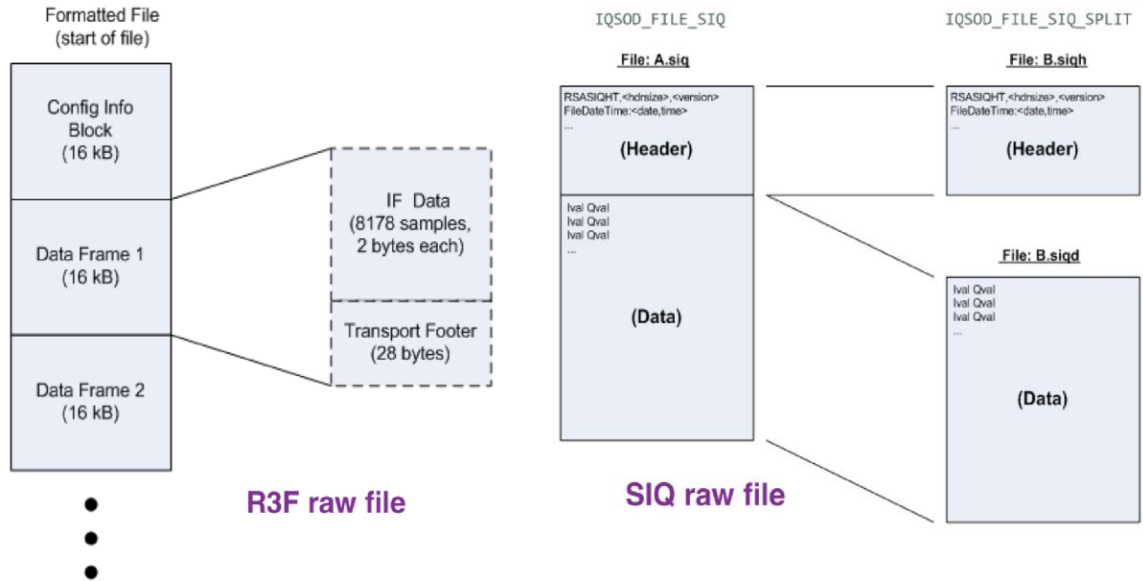

Fig. 9. .r3f and .siq file for data storage (Tektronix, 2017)

Fig. 10. .r3f and .siq file manager for data decompression if necessary.

As the RSA has the sampling rate of 112MHz and its signal sampling bandwidth of 40MHz, For each complete PD scanning process, the system will iteratively set RSA's IF modulation carrier frequency from 100MHz to 2500MHz (i.e., one iteration), with an incremental step of 40MHz, to cover the PD signal frequency of interest (Fig. 11.). In the end of each iteration (i.e., 61 samples in 60 iterations of 600 milli-seconds), the system will stitch the power-spectrum diagram of each sample frame for a complete diagram, and remove I/Q data that does not include any PD signals. If any PD signals are captured, the corresponding I/Q data and power-spectrum diagrams will be kept and named according to the timestamp and detected PD frequency. In this way, users can easily observe and trace back the records of PD occurrences, and the system storage can be more efficiently utilized.

## C. Remote Access to Continuous Real-Time PD Monitoring

With the expansion of the widely spread high-speed rail network, internet of things (IoT) technology should be added to local PD detection and monitoring systems at each switchgear/transformer station for the remote accessibility and potentially instant responsiveness to any alert messages it may generate. To achieve this goal, the google cloud service (e.g., Google Drive, Firebase, etc.) is employed and a smartphone application (APP) is developed. The local PD DAS will exploit Rclone (Craig-Wood, 2014) to set up a new cloud storage node, and continuously synchronize the newly detected PD I/Q data and full power-spectrum diagram to the cloud storage at the end of each PD scanning iteration. On the other hand, when the user activates the APP, it will periodically request data synchronization with the cloud server via Google drive API, so that the user can see the updated PD monitoring on their fingertips. In addition, whenever the cloud receives new files, it sends out a notification email to alert the user for newly detected PD events. The overall diagram of this cloud-based remote PD monitoring can be illustrated as Fig. 12.

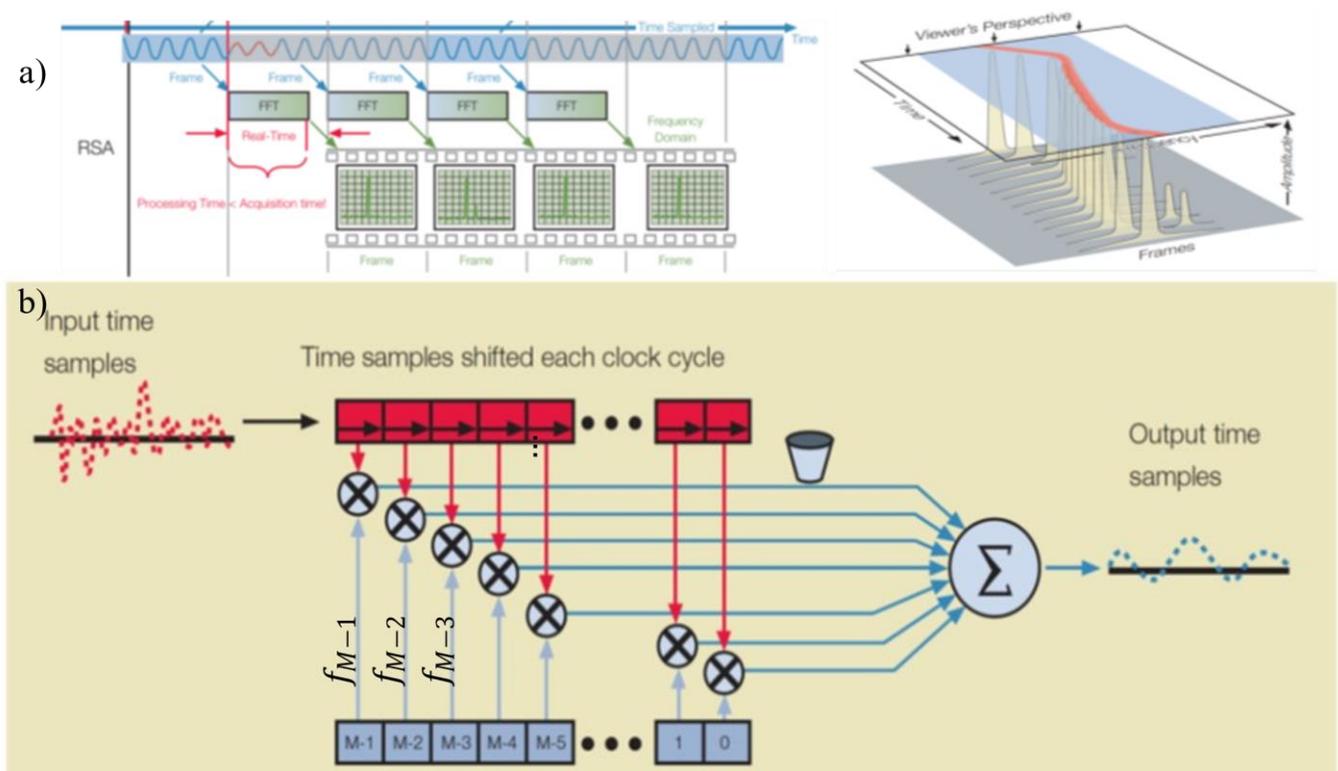

Fig. 11. a) Iterative sampling and spectrum analysis. b) Iteratively scanning to cover the entire PD spectrum.

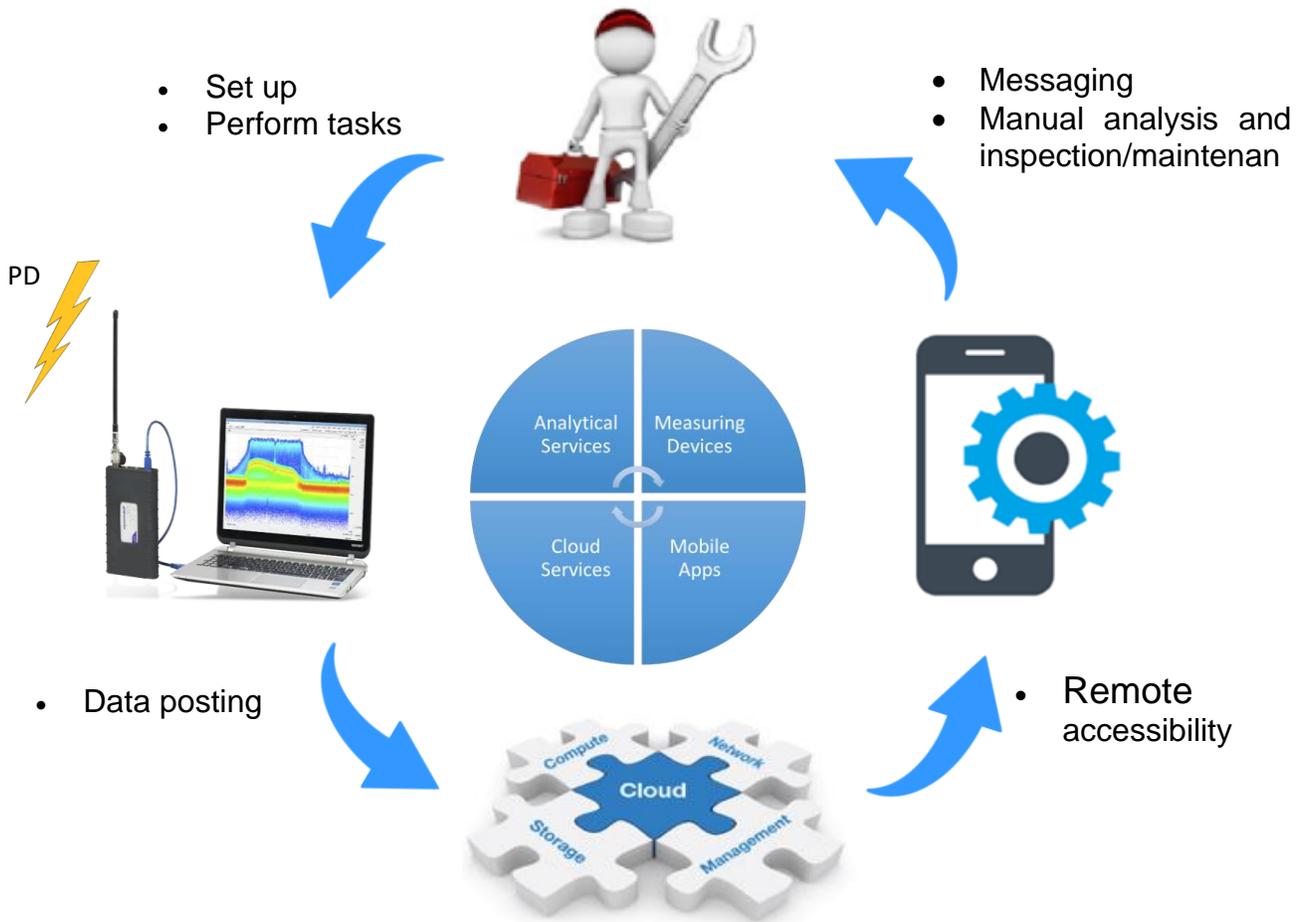

Fig. 12. IoT-cloud-based remote real-time PD monitoring system diagram

## III. SYSTEM EVALUATIONS

### A. Lab Environment and Hardware Setup

In this chapter, the system evaluation results are presented. The system has been tested under the lab environment (Fig. 13.), by applying a configurable arbitrary waveform generator (AWG) with a broad bandwidth of 2GHz (BNC, 2021). The AWG output is connected with a 50 resistive load, which is parallel connected with a 2G bandwidth oscilloscope (Tektronix, 2020) with a spectrum analyzer to measure the actual signal waveform, amplitude and frequency that is applied onto the resistor. When the AWG signal passes through the resistor, the circuit generates RF signals as a simulation of PD signals within 100~2500MHz range. Key AWG and oscilloscope specifications are listed in TABLE III. Numerical test scenarios have proceeded, and here we illustrate two representative cases, single signal and dual signals to prove the concept.

TABLE III Key specification of the test equipment in the system evaluation

| Equipment | Manufacturer | Model | Bandwidth |
|---|---|---|---|
| AWG | Berkeley Nucleonics Corp. (BNC) | 685 | 2 GHz |
| Oscilloscope | Tektronix | 5 Series MSO | 2 GHz |

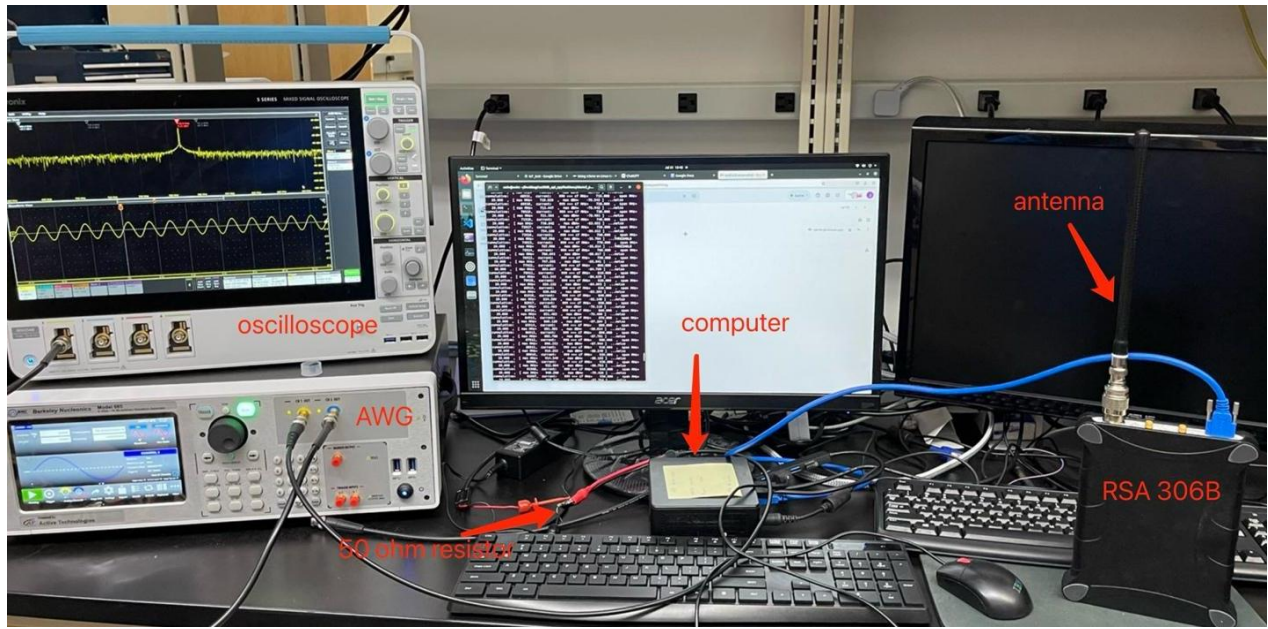

Fig. 13. Lab test setup for PD DAS (including antenna, RSA306B, a portable high-performance computer/embedded system), an AWG and 50-ohm resistor (i.e., simulator of PD source), and an oscilloscope.

*B. Single RF/PD Signal Capturing*

We connected the 50 resistor with AWG via BNC-alligator cable, and placed the PD DAS with its antenna about one foot away from the resistor. Once the PD DAS is turned on and initiated with the program to scan RF/PD signals in the air (Fig. 14.), we set up the AWG to generate a 315 MHz sine waveform of 1V peak-to-peak. Then, the DAS captures the RF signals that exceed the preset threshold (Fig. 15.), and generates one spreadsheet (i.e., .csv file) of signal value in time-domain (i.e., I/Q ADC values), and a power-spectrum diagram in frequency domain (Fig. 16.a and b). In the end of that particular scanning iteration, a complete power-spectrum diagram is drawn from 100MHz to 2500MHz (Fig. 16.c). Another test results of 768 MHz RF/PD signal are demonstrated in Fig. 17. One can see that although there exist some relatively low frequency noises, most less than 1MHz and -60dB, RF/PD signals of interest can be clearly captured without much interference. In addition, the IF modulation can be observed in Fig. 16.a and especially in Fig. 17.a, because of the higher IF carrier frequency.

Fig. 14. Loading program on the PD DAS and getting ready for the PD detection and monitoring.

![Fig. 15 terminal output]

Fig. 15. Capturing signals above the preset power threshold.

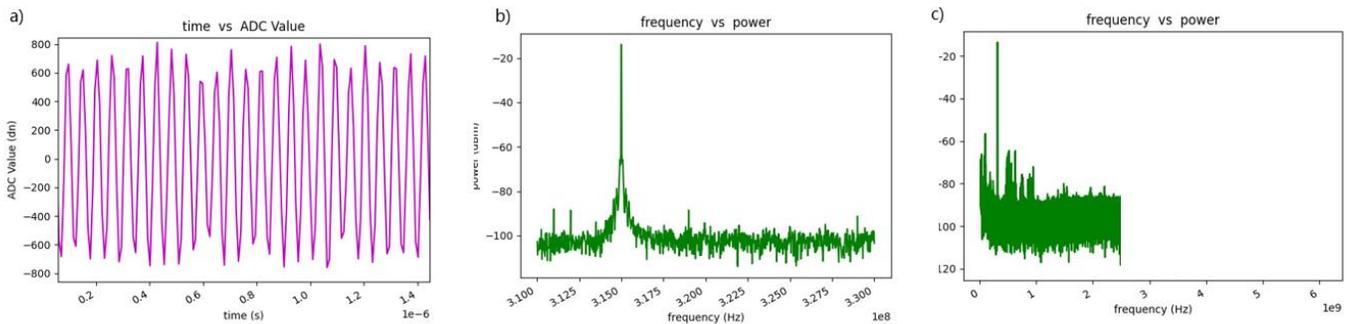

Fig. 16. Capturing 315MHz RF/PD signals a) I/Q ADC values in time domain, b) impulse at 315MHz in frequency domain for one frame, c) impulse at 315MHz in complete scanning iteration

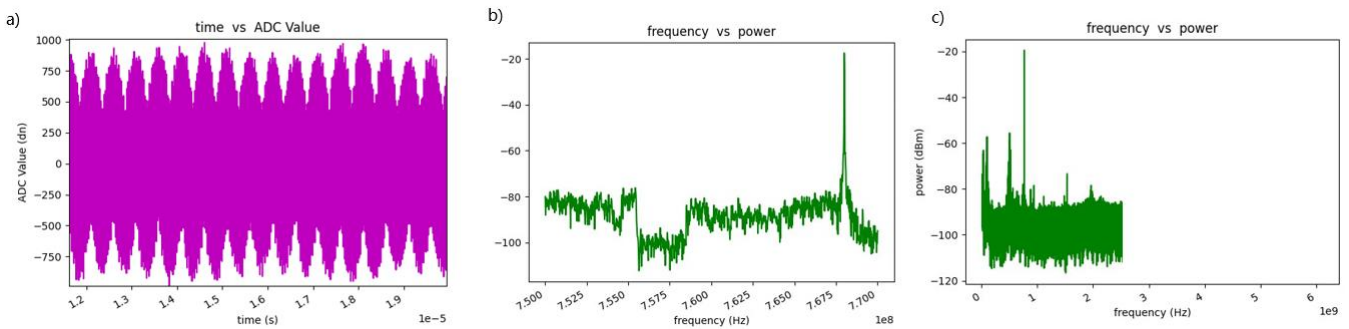

Fig. 17. Capturing 768MHz RF/PD signals a) I/Q ADC values in time domain, b) impulse at 768MHz in frequency domain for one frame, c) impulse at 768MHz in complete scanning iteration

## C. Dual/Multiple RF/PD Signal Capturing

We also have run the dual RF/PD detection test to confirm that if multiple PD events occur approximately at the same time (i.e., in one scanning iteration), the PD DAS can capture both of them. We first create a Bluetooth signal (i.e., about 2400 MHz) by turning on two Bluetooth devices and transmitting large files from one device to the other. Meanwhile, we create another 768 MHz signal with AWG as aforementioned, and turn on the PD DAS. It is worth noting that we have to adjust the RF/PD detection threshold to -70dB so as to catch the low power Bluetooth signal. As depicted in Fig. 18, one can see that the DAS can clearly capture both 768MHz and the Bluetooth signals.

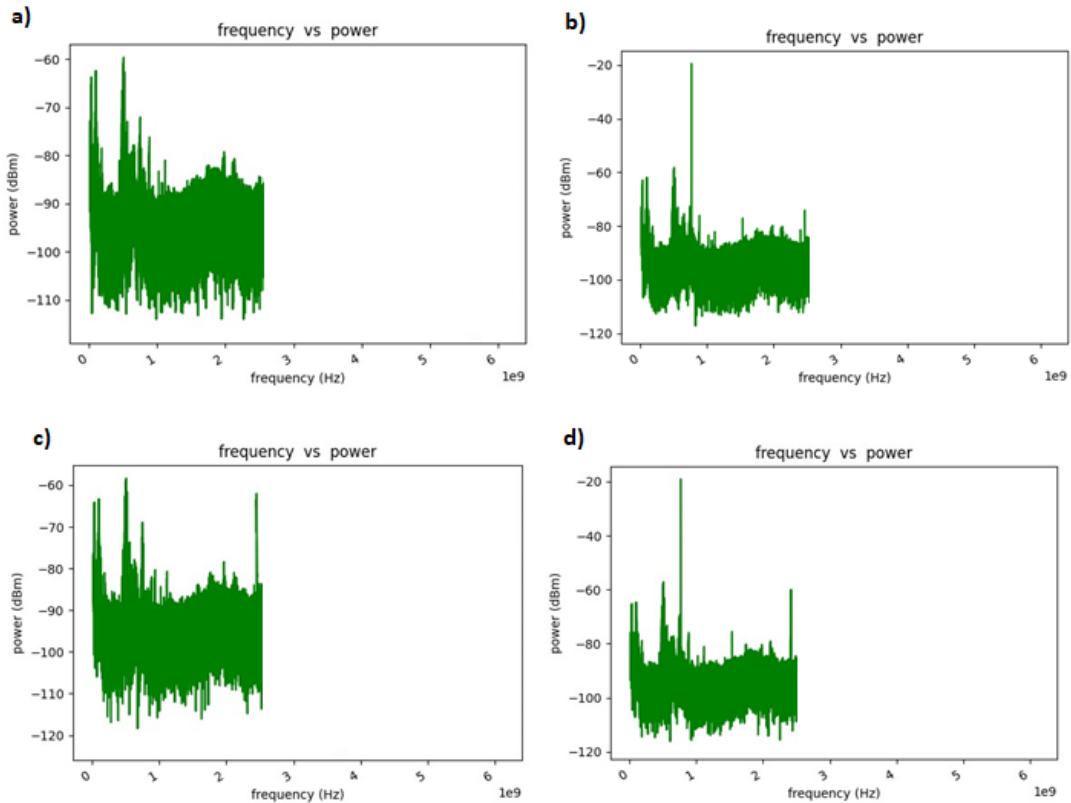

Fig. 18. a) No test signal generated (only noises), b) only with a 768MHz generated by AWG, c) only with a Bluetooth signal at about 2400MHz, d) with both 768MHz and Bluetooth signals in the air.

*D. Cloud-Based Remote Monitoring*

The cloud service is established on the DAS via Google drive and Rclone, so that any files created and edited within this particular Google drive folder will be automatically synchronized in all terminals that use the same Google drive account. Once the user logs in with proper credentials (Fig. 19. a) on a remote device (e.g., a smartphone), one can see the updated I/Q data and spectrum diagrams (Fig. 19. b), and corresponding file attribute details (Fig. 19. c). In addition to the real-time synchronization from the local PD DAS to the cloud and granting remote accessibility, the cloud also sends a notification email to the user whenever any new PD signals are detected (Fig. 20.). In that case, personnel who are in charge of monitoring the switchgears and transformers in high-speed rail can be alerted at an early stage to minimize the potential damage and maintenance cost PD may cause.

IV. CONCLUSION

This study has successfully developed an advanced platform for the non-disruptive, continuous, real-time detection and monitoring of partial discharges (PD) in switchgears and transformers used within high-speed rail infrastructure. The system is designed to iteratively scan and capture PD signals within a frequency range of 100MHz to 2500MHz, ensuring comprehensive coverage and detection of PD activities.

In addition to its robust detection capabilities, the system is seamlessly integrated with cloud services, enabling remote accessibility to real-time PD data and analysis results. This feature facilitates prompt decision-making and timely interventions by maintenance teams, significantly enhancing the overall reliability and safety of high-speed rail operations.

Extensive laboratory evaluations have been conducted to assess the system's performance. These evaluations demonstrate the system's effectiveness and proficiency in automated real-time PD detection and long-term monitoring. The results indicate that the platform not only meets but exceeds the operational requirements, ensuring that potential faults are identified and addressed promptly before they can escalate into major issues.

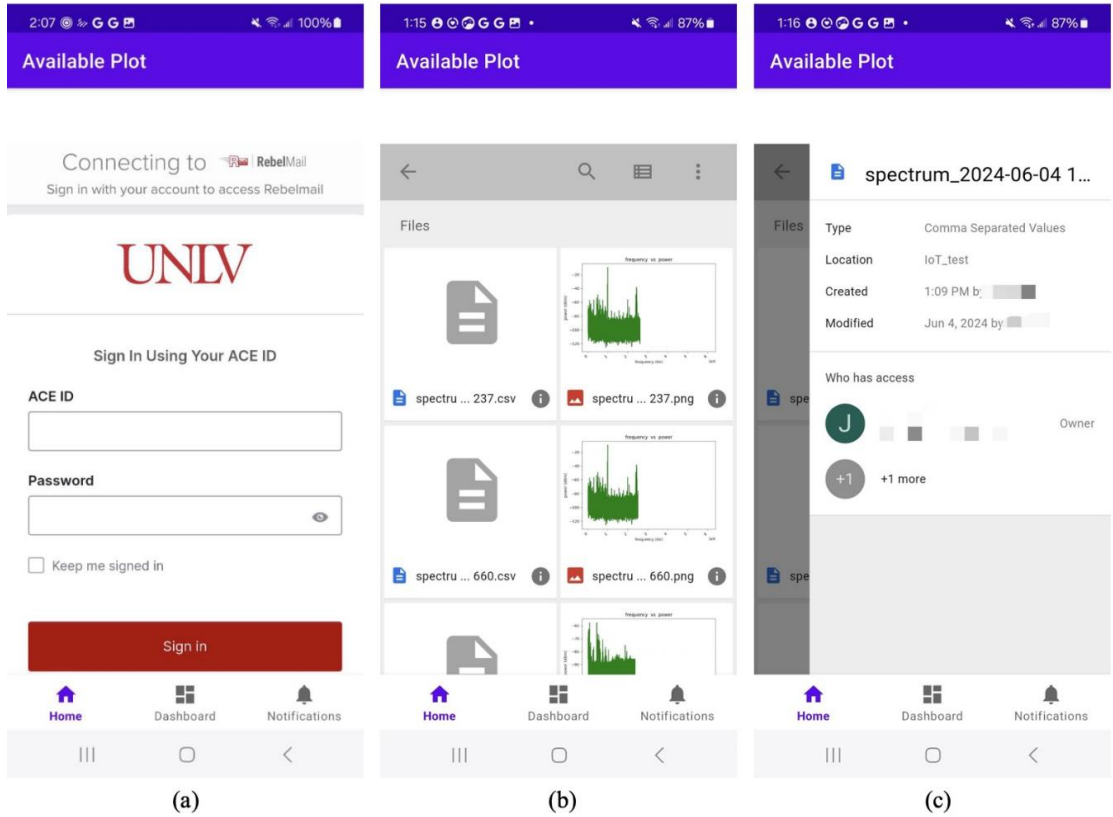

Fig. 19. a) APP login, b) APP access to files, c) File attribute details

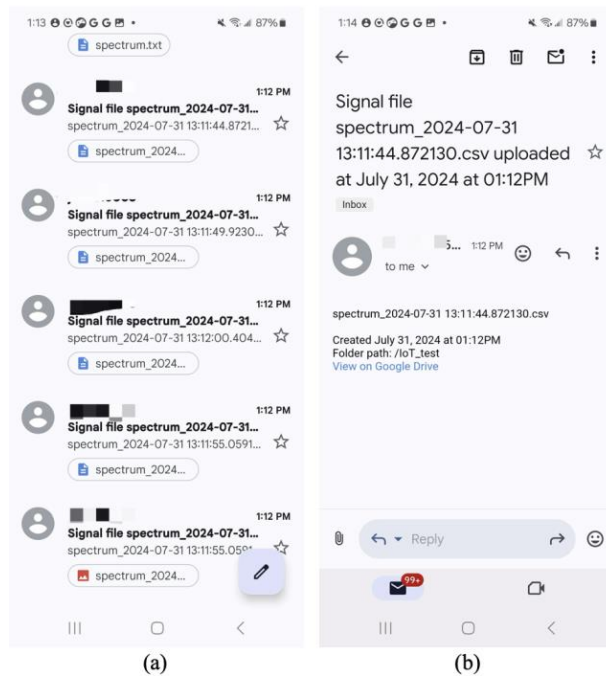

Fig. 20. Email notification of detecting new PD signals

The successful implementation of this platform marks a significant advancement in the maintenance and monitoring of high-voltage electrical infrastructure in high-speed rail systems. By leveraging real-time data and remote monitoring capabilities, this system represents a proactive approach to infrastructure management, reducing the risk of unexpected failures and enhancing the overall efficiency and safety of high-speed rail operations.

## V.     ACKNOWLEDGMENT

This study was conducted with the support from the USDOT Tier 1 University Transportation Center on Railroad Sustainability and Durability with the grant no. of 69A3551747132.